\documentclass[doublecol]{epl2}
\usepackage{amsmath, bm, physics, amssymb}
\definecolor{linkcolor}{rgb}{0,0,0.6} 
\usepackage[pdftex,colorlinks=true,
	pdfstartview = FitV,
	linkcolor    = linkcolor,
	citecolor    = linkcolor,
	urlcolor     = linkcolor,
	hyperindex   = true,
	hyperfigures = false]{hyperref}
\usepackage{soul}
	
\newcommand{\bx}{{\boldsymbol{x}}}
\newcommand{\br}{{\boldsymbol{r}}}

\newcommand{\bv}{{\boldsymbol{v}}}

\newcommand{\bLambda}{\boldsymbol{\Lambda}}

\title{Towards a liquid-state theory for active matter}

\author{Yuting Irene Li \inst{1} \and Rosalba Garcia-Millan\inst{1, 3} \and Michael E. Cates\inst{1} \and \'Etienne Fodor \inst{2}}
\shortauthor{Y. I. Li \etal}

\institute{   
	\inst{1} DAMTP, Centre for Mathematical Sciences, University of Cambridge, Wilberforce Road, Cambridge CB3 0WA, UK
	\\                 
  \inst{2} Department of Physics and Materials Science, University of Luxembourg, L-1511 Luxembourg, Luxembourg
  \\
	\inst{3} St John's College, University of Cambridge, Cambridge CB2 1TP, UK
}


\abstract{
In equilibrium, the collective behaviour of particles interacting via steep, short-ranged potentials is well captured by the virial expansion of the free energy at low density. Here, we extend this approach beyond equilibrium to the case of active matter with self-propelled particles. Given that active systems do not admit any free-energy description in general, our aim is to build the dynamics of the coarse-grained density from first principles without any equilibrium assumption. Starting from microscopic equations of motion, we obtain the hierarchy of density correlations, which we close with an ansatz for the two-point density valid in the dilute regime at small activity. This closure yields the nonlinear dynamics of the one-point density, with hydrodynamic coefficients depending explicitly on microscopic interactions, by analogy with the equilibrium virial expansion. This dynamics admits a spinodal instability for purely repulsive interactions, a signature of motility-induced phase separation. Therefore, although our approach should be restricted to dilute, weakly-active systems a priori, it actually captures the features of a broader class of active matter.
} 

\begin{document}

\maketitle


Active matter is a wide category of systems out of equilibrium, where a net flow of energy takes place at the local, individual level~\cite{Marchetti2013, Wijland2022, Jack2022}. Examples are swimming bacteria~\cite{ElgetiETAL:2015} and active emulsions~\cite{weber_review:2019}, amongst others. The realm of motile active matter includes interacting, many-particle systems, where particles move persistently. Different theoretical approaches have been proposed to describe such systems. They include (i) particle-based models, such as run-and-tumble particles (RTPs)~\cite{tailleur:2008}, active Brownian particles (ABPs)~\cite{filyPRL:2012}, and active Ornstein-Uhlenbeck particles (AOUPs)~\cite{FodorPRL, MartinPRE}, (ii) microscopic field theories~\cite{Garcia-MillanPruessner:2021, PruessnerGarcia-Millan:2022}, and (iii) coarse-grained field theories~\cite{Nardini2017, JulicherETAL:2018, tjhung:2018}.

Self-propelled particles tend to slow down at high densities, due to either biochemical reasons or steric repulsions. In contrast with passively diffusing particles, this slowdown in turn increases the local density, creating a positive feedback loop that can result in a phase separation between a dense and a dilute phase, known as motility-induced phase separation (MIPS)~\cite{mips}. To predict analytically the emergence of MIPS, previous works have relied on a local mean-field approximation to replace microscopic interaction with density-dependent motility~\cite{stenhammarPRL}, and also on an adiabatic elimination of orientational degree of freedom~\cite{mips}. Other studies have developed some aspects of a liquid-state theory for active matter with pairwise interactions. This is done, for instance, by expressing thermodynamic observables, such as pressure~\cite{Brady2014, Marchetti2014, pressure} and dissipation~\cite{dissipation, Nemoto2020}, in terms of correlations between hydrodynamic fields, typically density and polarization. Also, exact results for density correlations have been obtained at infinite dimensions~\cite{Pirey2019, Pirey2021}.

The success of equilibrium thermodynamics largely stems from the ability to detect phase transitions by analyzing the free-energy. In the dilute regime, the virial expansion provides an approximate expression of the free-energy that averages the effect of steep, short-ranged pairwise interactions~\cite{KardarParticles}. For passive Brownian particles (PBPs) interacting with an isotropic pairwise potential $\sum_{i,j<i}\Psi(|{\bf x}_i-{\bf x}_j|)$ at temperature $T$, the free energy per unit volume is given in terms of the uniform density $\rho$ as
\begin{equation}
\begin{split}
	f_\mathrm{virial} (\rho) / T&= \rho \log \rho + B(T) \rho^2  + O(\rho^3),
	\\
	B &= \frac{1}{2} \int \dd \br \left ( 1 - e^{-\Psi (r) /T} \right ). 
\end{split} 
\label{eq:1}
\end{equation}
Here $B$ is commonly referred to as the second virial coefficient. The major success of the virial expansion lies in predicting the onset of liquid-gas phase separation in particle systems with a combination of strongly repulsive short-ranged forces and weakly attractive long-range forces, such as the Lennard-Jones potential. As temperature increases, the free energy, as a function of density $\rho$, goes from convex ($B>0$) to concave $(B<0)$: It yields a phase transition from homogeneous to non-homogeneous density profiles, with a separation between dilute and dense phases. Importantly, the virial expansion holds for a generic microscopic potential (excluding those of such long range that the excess free energy is not analytic in density).

In classical thermodynamics, the virial expansion is often derived from the equilibrium partition function~\cite{KardarParticles}. Interestingly, it is also possible to arrive at approximate expressions of free energy by truncating the hierarchy of density correlations, known as the Bogoliubov–Born–Green–Kirkwood–Yvon (BBGKY) hierarchy~\cite{hansen_mcdonald}. For instance, using an ansatz for the two-particle density, informed by the steady-state solution of the two-body problem~\cite{GrosbergPRE}, the dynamics of the one-body density follows a gradient flow with respect to $f_{\rm virial}$, as shown in the SM. This type of derivation can potentially be extended to a large class of nonequilibrum systems, where the dynamics does not derive from any free energy {\it a priori}. Such an extension requires proposing an ansatz for the two-particle density, which should be appropriate to the specific nonequilibrium system at hand.

Interestingly, the two-body probability distribution can be approached perturbatively for AOUPs, where the persistence time is the small parameter in natural units~\cite{FodorPRL, MartinPRE}. To first order, this solution already predicts that repulsive interaction yields effective attraction. This suggests that the onset of MIPS can likewise already be captured by approximating the dynamics of density at this order. Therefore, inspired by the equilibrium virial expansion, it is tempting to explore whether the density dynamics already contains any spinodal instability for dilute active systems at weak persistence. Note that, although previous work refer to ``active virial'' as an expansion of pressure~\cite{virial_pressure, virial_pressure2}, here we are instead interested in deriving {\em dynamical equations of motion for the density}. Indeed, at variance with equilibrium, phase transitions in active systems are usually not thermodynamically controlled by any equation of state, but they can still be directly detected as instability in the dynamics. Once the density equations are derived for the density, either by hydrodynamic closures \cite{Wittkowski:NJP:2017} or from symmetry arguments \cite{mips}, the onsets of instability are typically found by linear perturbations around the uniform state, as illustrated in several recent works \cite{morin:natphys:2017, Wagner-Baskaran:JStat:2017, Peshkov-Chate:PRL:2012, Wittkowski:NJP:2017}.

In this Letter, we start with the particle-based description of AOUPs, and consider their steady-state probability distribution. Our roadmap is essentially a ``small density-small persistence'' bottom-up derivation of the density dynamics. To this end, we first introduce the corresponding hierarchy of density correlations. Inspired by equilibrium thermodynamics~\cite{hansen_mcdonald, GrosbergPRE}, we then close the hierarchy to arrive at a nonlinear dynamics for the one-point density. It allows us to identify the MIPS spinodal instability for \textit{any} soft microscopic repulsive interactions. An advantage of our approach is that does not impose an equilibrium mapping \textit{a priori} (see~\cite{FodorPRL, MartinPRE} for a discussion of closures to the AOUP equations that do so), instead retaining the non-equilibrium structure of the microscopic theory. A disadvantage is that extensions to higher order in density would involved increasingly complicated calculations of higher virial coefficients, which we do not attempt here.


{\bf Interacting AOUPs: Joint distribution function}~--~We consider a system of $N$ AOUPs at positions $\bx_i$ interacting with pairwise potential $U$. The particles are subject to stochastic self-propulsion velocities ${\bv}_i$ with persistence time $\tau$ and diffusivity $D$~\cite{FodorPRL, MartinPRE}: 
\begin{equation}
\begin{split} 
	\dot\bx_i  &= \bv_i - \partial_{\bx_i} U ,
	\quad
	U = \sum_{i=1}^N \sum_{j<i} \Psi ( r_{ij} ) ,
	\\
	\tau\dot \bv_i &= - \bv_i + \sqrt{ 2 D} \bLambda_i ,
\end{split} 
\label{eq:eom}
\end{equation}
where $r_{ij} = | \bx_i - \bx_j |$, and we have set the mobility to unity. Here $\bLambda_i$ is a Gaussian white noise, with correlation $\expval{ (\bLambda_i)_\alpha(t) (\bLambda_j)_\beta(0) } = \delta_{ij} \delta_{\alpha\beta} \delta(t)$, where latin and Greek letters respectively denote particle index and spatial coordinates. It follows that $\bv_i$ is a colored Gaussian noise, with correlation $\expval{ (\bv_i)_\alpha(t) (\bv_j)_\beta(0) } = \delta_{ij} \delta_{\alpha\beta} (D/\tau) e^{-|t|/\tau}$.

In the limit of vanishing persistence ($\tau\to0$), the correlation of $\bv_i$ becomes white, so that the system reduces to a set of overdamped PBPs. Note that, at finite $\tau$, if the potential term $-\partial_{\bx_i} U$ was in the rhs of the $\bv_i$-equation (instead of the $\bx_i$-equation), the system would represent underdamped PBPs in equilibrium. For this reason, overdamped AOUPs and underdamped PBPs coincide in the noninteracting limit.

Following standard procedures~\cite{risken}, already widely used in active matter (see e.g.~\cite{Saintillan:review:2018, Bialke:2015}), the Fokker-Planck equation of the joint probability distribution $p_N(\bx_1, \bv_1, \bx_1, \bv_2, \dots ,\bx_N, \bv_N, t)$ associated with eq.~\eqref{eq:eom} is
\begin{equation}
\begin{split} 
	\partial_t p_N  &= \sum_{i=1}^N \Big [ \mathcal{L}_{\mathrm{f},i} p_N + \partial_{\bx_i}  \cdot \left (  p_N \, \partial_{\bx_i} U  \right ) \Big ] ,
	\\
	\mathcal{L}_{\mathrm{f},i}  &= \frac{D}{\tau^2 } \partial_{\bv_i}^2 + \left ( \frac{1}{\tau} \partial_{\bv_i} - \partial_{\bx_i} \right ) \cdot \bv_i, 
	\end{split} 
\label{eq:fp}
\end{equation}
where $\mathcal{L}_{\mathrm{f},i}$ is the non-interacting Fokker-Planck operator governing the free-particle motion of the $i$-th particle. Although it cannot be solved exactly, its steady state can be computed to lowest orders in the persistence time $\tau$ in a similar manner to~\cite{MartinPRE, FodorPRL}. Such a perturbative calculation can be performed in arbitrary dimension, though we now restrict ourselves to 1D for simplicity, as a proof of principle. After scaling units as $v \to v \sqrt{\tau/D}$ , $x \to x/\sqrt{\tau D}$, $t \to t/\tau$, and $\Psi \to \Psi / D$, we find the many-body steady state probability [see SM for details], 
\begin{equation}
\begin{split} 
	p_N &\propto e^{ - U - \frac{ v_1^2 + v_2^2 }{2} } \bigg [ 1 + \sqrt{ \tau} \sum_{i=1}^N v_i \partial_{x_i} U
	\\ 
	&\quad + \tau \sum_{i=1}^N \left ( \frac{v_i^2}{2} \left ((\partial_{x_i} U)^2 - \partial_{x_i}^2 U  \right ) - (\partial_{x_i} U) ^2 + \frac{3}{2} \partial_{x_i}^2 U \right )
	\\
	&\quad + o(\tau) \bigg ] .
\end{split} 
\label{eq:fps}
\end{equation}
Note that the steady state probability here is given in $(x,v)$ space, whereas~\cite{FodorPRL} gives the corresponding probability in $(x,\dot x)$ space.


{\bf Nonequilibrium density correlations: Hierarchy of equations}~--~Although the joint distribution function $p_N$ in eq.~\eqref{eq:fps} contains all information about the steady state density, it is generally difficult to predict the emergence of phase transitions from the perspective of the full phase space. Instead, our aim is to obtain a reduced description of the systel in terms of density correlations. The $n$-particle density $p_n$ is found by marginalising the joint distribution function $p_N$ as
\begin{equation} 
\begin{split}
	&p_n ( \bx_1, \bv_1, \dots, \bx_n, \bv_n, t )
	\\
  & \, = P_{N,n} \int p_N (\bx_1, \bv_1, \dots , \bx_N, \bv_N, t ) \prod_{j=n+1}^N \dd \bx_j\dd \bv_j ,
\end{split} 
\end{equation} 
where $P_{N,n} = N ! / (N-n)!$ is the permutation coefficient. Thereafter, for simplicity, we use the shorthand notation $p_n = p_n ( \bx_1, \bv_1, \dots, \bx_n, \bv_n, t )$ for arbitrary $n$, where the dependence on positions, self-propulsion velocities, and time will be omitted.  Integrating eq.~\eqref{eq:fp} over the positions and velocities of all particles but the first $n$ yields a hierarchy of equations, each depending on the $n+1$-particle density $p_{n+1}$, analoguous to the BBGKY hierarchy of PBPs~\cite{hansen_mcdonald}. The first such equation ($n=1$) is as follows, 
\begin{equation}
\begin{split} 
	\partial_t p_1 &= \mathcal{L}_{\mathrm{f},1} p_1 + F_1 [ p_2] ,
	\\
	F_1 [p_2] &= \partial_{\bx_1} \cdot \int p_2 \, \partial_{\bx_1} \Psi (r_{12}) \,\dd \bx_2 \dd \bv_2. 
\end{split} 
\label{eq:p1}
\end{equation} 
The first term $\mathcal{L}_{\mathrm{f},1} p_1$ corresponds to the free single-particle dynamics, whereas the second term $F_1 [ p_2]$ represents the effects of pairwise interaction between one particle and all the others. Similarly, the second equation of the hierarchy is, 
\begin{equation}
\begin{split} 
	\partial_t p_2 &= ( \mathcal{L}_{\mathrm{f},1} + \mathcal{L}_{\mathrm{f},2} + \mathcal{L}_\mathrm{int}  ) p_2 + F_2 [p_3] ,
	\\
	\mathcal{L}_\mathrm{int} &= \sum_{i=1,2} \partial_{\bx_i} \cdot \Big[ \partial_{\bx_i} \Psi ( r_{12} ) \Big] ,
	\\
	F_2 [p_3] &= \sum_{i=1,2} \partial_{\bx_i} \cdot \int p_3 \, \partial_{\bx_i} \Psi \left ( r_{i3} \right ) \, \dd \bx_3 \dd \bv_3  .
\end{split} 
\label{eq:p2p3}
\end{equation} 
Here $\mathcal{L}_\mathrm{int}$ governs the pairwise interactions between any two particles, and  $F_2 [p_3]$ represents the interactions between either of the first two particles and all the other particles. As we will show now, this is a higher-order term in the average number density $\rho = N/V$ compared to the rest, and hence can be neglected in the dilute limit.

To determine the scaling of $p_n$  with respect to $\rho$, one can start with their normalisation properties~\cite{KardarParticles}
\begin{equation} 
	\int p_n \,\prod_{i=1}^n \dd \bx_i \dd \bv_i = P_{N,n} ,
\end{equation} 
from which follows $p_n \sim \rho^n$ for small $n$. Physically, this is indeed expected as $p_1 (\bx_1, \bv_1 ) $ is the probability density of finding $n$ particles at a given set of coordinates. With this scaling in mind, we estimate the scaling of each term in eq.~\eqref{eq:p2p3} as
\begin{equation}\label{eq:sc}
\begin{split}
	&( \partial_t - \mathcal{L}_{\mathrm{f},1} - \mathcal{L}_{\mathrm{f},2} - \underbrace{\mathcal{L}_\mathrm{int}}_{\sim 1/\tau_\mathrm{int}} )  \underbrace{p_2}_{\sim  \rho^2 }
	\\
	&\quad = \sum_{i=1,2} \partial_{\bx_i} \cdot \int \underbrace{p_3}_{\sim \rho^3} \, \underbrace{\partial_{\bx_i} \Psi ( r_{i3} )}_{ \sim 1/\tau_\mathrm{int}} \,\underbrace{ \dd\bx_3 }_{\sim (r_0)^d} \dd \bv_3  ,
\end{split} 
\end{equation} 
where $\tau_\mathrm{int}$ is the typical timescale of interaction, $r_0$ is the lengthscale of the potential, and $d$ denotes the sptial dimension. One can see that the rhs of eq.~\eqref{eq:sc} is small if $r_0^d \rho \ll 1$. Since $r_0^d$ is effectively the volume of a particle, this condition is satisfied if the volume fraction of particles in the system is low. Therefore, as expected from the analogy with the BBKY hierarchy~\cite{hansen_mcdonald}, the interaction of two particles with respect to others in the bath can be neglected at small volume fraction.


{\bf Closure of hierarchy: Insight from quasistatic approximation}~--~We now propose a closure of the hierarchy of equation for density correlations. At small volume fraction, neglecting the rhs in eq.~\eqref{eq:sc} yields
\begin{equation}\label{eq:p2}
	\partial_t p_2  = ( \mathcal{L}_{\mathrm{f},1} + \mathcal{L}_{\mathrm{f},2} + \mathcal{L}_\mathrm{int} ) \, p_2 .
\end{equation}
The dynamics in eq.~\eqref{eq:p2} is equivalent to the Fokker-Planck equation of two AOUPs interacting through the pairwise potential $\Psi(r_{12})$. Recall from eq.~\eqref{eq:fps} the N-particle stationary probability, calculated order by order in the persistence time $\tau$. The two-particle probability is hence, 
\begin{equation}
	p_{2,\mathrm{ss}}(r_{12},v_1,v_2) \propto e^{ - \Psi(r_{12}) - \frac{v_1^2 + v_2^2}{2} } \, g (r_{12}, v_1-v_2) ,
\end{equation}
where
\begin{equation} 
\begin{split} 
	g (r, w) &=  1 + \sqrt{\tau} w \Psi' + (\tau w^2/2) (  ( \Psi')^2 - \Psi '' )
	\\
	&\quad + \tau ( 3 \Psi'' - 2 (\Psi' )^2 ) + O(\tau^{3/2}) ,
\end{split}
\label{eq:Trans}
\end{equation} 
and $\Psi' = \dd\Psi/\dd r$ [see SM]. In the quasistatic limit, where the interaction timescale is much shorter than the persistence ($\tau_\mathrm{int} \ll \tau $), we assume that every pair of particles reaches steady state. Inspired by previous works~\cite{GrosbergPRE, IlkerPRR}, this assumption motivates the following ansatz to close the hierarchy of density correlations:
\begin{equation}\label{eq:ansatz}
\begin{split} 
	p_2 (x_1, v_1, x_2, v_2, t) &= p_1 (x_1, v_1, t) \, p_1 ( x_2, v_2, t )
	\\
	&\quad\times e^{- \Psi(r_{12})} g (r_{12}, v_1-v_2) .
\end{split} 
\end{equation} 
There is evidence that the above ansatz works well for strong short-ranged interactions in the equilibrium case [see SM]. We assume it would also work in the non-equilibrium case. Indeed, when the interparticle distance is larger than the interaction range ($r_{12}>r_0$), the two-point function $p_2$ reduces to the product of one-point functions $p_1$, as expected from a mean-field approach for noninteracting systems. The second line of eq.~\eqref{eq:ansatz} accounts for corrections due to interactions. For repulsive potential, the factor $e^{-\Psi}$ captures the reduction of the two-point density caused by the interaction, as expected from equilibrium, whereas $g (r_{12}, v_1-v_2)$ encapsulates nonequilibrium effects. We highlight that the ansatz naturally encodes non-trivial velocity-velocity correlations that are absent in equilibrium, similar to the findings of \cite{Marconi:2021}. The full consequences of such correlations are beyond the scope of this letter. Note, however, that related velocity correlations are known in several other nonequilibrium contexts including but not limited to granular matter~\cite{bertin:review:2017, Brilliantov:book}.

Substituting the ansatz from eq.~\eqref{eq:ansatz} into eq.~\eqref{eq:p1}, we arrive at the following dynamics for the one-particle density:
\begin{equation}
\begin{split} 
	(\partial_t - & \mathcal{L}_\mathrm{f}) \, p_1 (x, v, t)
	\\
	&= \tau \partial_x \bigg[ p_1(x,v,t) \int {\cal L}_{\rm v} p_1(x,v-w,t) \dd w \bigg] .
\end{split} 
\label{eq:full_eq}
\end{equation}
Here, $\mathcal{L}_\mathrm{f}$ is the free Fokker-Planck operator in scaled units, and the operator ${\cal L}_{\rm v}$ effectively takes into account interactions:
\begin{equation}
\begin{split} 
	\mathcal{L}_\mathrm{f} &= \partial_v^2 +\left ( \partial_v  -  \sqrt{ \tau} \partial_x \right ) v ,
	\\
	{\cal L}_{\rm v} &= \int \Psi'(r) e^{-\Psi(r)} g(r, w) e^{-r\partial_x} \dd r ,
\end{split}
\label{eq:op}
\end{equation}
where we have introduced the translation operator $e^{- r \partial_x}$, which shifts the position of the function acted upon as $e^{- r \partial_x} p_1(x, v) = p_1(x-r, v)$. Hence, ${\cal L}_{\rm v}$ corresponds to an infinite series of the gradient $\partial_x$, with series coefficients set by the microscopic potential. Equation~\eqref{eq:full_eq} is the central result of this Letter: It provides the dynamics of the one-particle density in a closed form for small average density. Importantly, this closed form depends explicitly on the details of the pairwise potential $\Psi$.

To obtain a more explicit expression of ${\cal L}_{\rm v}$, we next substitute our perturbative result for $g$ [eq.~\eqref{eq:Trans}] into the definition of ${\cal L}_{\rm v}$ [eq.~\eqref{eq:op}], yielding
\begin{equation}\label{eq:Lv}
	{\cal L}_{\rm v} = 2 w {\cal L}_{\rm a} + 2 {\cal L}_{\rm b} \partial_x + w^2 {\cal L}_{\rm c} \partial_x ,
\end{equation}
where, after integration by parts, we get
\begin{equation}\label{eq:Labc}
\begin{split} 
	{\cal L}_{\rm a} &=  \sqrt{\tau} \int_0^\infty \dd r (\Psi' )^2 \, e^{-\Psi} e^{ - r \partial_x } ,
	\\
	{\cal L}_{\rm b} &= -\int_0^\infty \dd r f_0(r) e^{ - r \partial_x } ,
	\\
	{\cal L}_{\rm c} &= - \int_0^\infty \dd r \, f_1(r)  e^{ - r \partial_x } ,
\end{split} 
\end{equation}
and
\begin{equation}
\begin{split} 
	f_0 (s) &=  \int_s^\infty \dd r \, \Psi' e^{-\Psi} \left [ 1 -  \tau \left ( 2 ( \Psi' )^2  - 3  \Psi'' \right)   \right ] ,
	\\
	f_1(s) &= \tau \int_s^\infty \dd r \,  \Psi'  \left ( (\Psi')^2 - \Psi'' \right ) e^{-\Psi} .
\end{split} 
\end{equation}
The integrands featuring in the definition of the operators $\{{\cal L}_{\rm a}, {\cal L}_{\rm b}, {\cal L}_{\rm c}\}$ [eq.~\eqref{eq:Labc}] are even with respect to $r$, provided that $\Psi(r) = \Psi(-r)$. It follows that these operators can be expressed as a series of $\partial_x^2$.

The functions $f_0$ and $f_1$ are analoguous to the Mayer function which appears in the equilibrium virial expansion~\cite{hansen_mcdonald}. Indeed, in the equilibrium limit ($\tau \rightarrow 0$), the only surviving term in ${\cal L}_{\rm v}$ stems from the leading order in $f_0$, in agreement with the derivation for equilibrium overdamped PBPs [see SM]. This term is responsible for the $B$ coefficient in eq.~\eqref{eq:1} when that result is derived dynamically for equilibrium systems. In that respect, eq.~\eqref{eq:Lv} can be regarded as a direct generalization of the equilibrium virial expansion to the AOUP setting. Importantly, eq.~\eqref{eq:Lv} shows that activity produces additional velocity-dependent terms that are absent in equilibrium. In what follows, we analyze in detail the corresponding dynamics, in search for the onset of instability as a signature of MIPS.


{\bf Linear stability analysis: Eigenvalue problem}~--~The stationary solution of eq.~\eqref{eq:full_eq} is given by $ p_1^{(0)}(v) = (\rho /\sqrt{2 \pi}) e^{-v^2/2}$. This solution corresponds to a uniform density for the position and a Gaussian distribution for the self-propulsion. In addition, as we will see shortly, this is also the ground state of the free operator. In the following, we expand perturbatively around $p_1^{(0)}$ to find its instability regions in parameter space: If the uniform state is not stable, then we argue that the system undergoes spinodal decomposition via a MIPS mechanism.

We consider $p_1 (x, v, t) =  p_1^{(0)} (v) + \varepsilon(x, v,t)$, with $\varepsilon$ a small perturbation about the ground state $p_1^{(0)}$. Expanding eq.~\eqref{eq:full_eq} to linear order in $\varepsilon$, we get
\begin{equation}
	(\partial_t - \bar{\mathcal{L}}_\mathrm{f}) \,\varepsilon(x,v,t) = \tau p_1^{(0)}(v) \int {\cal L}_{\rm v} \partial_x \varepsilon(x, v-w, t) \dd w ,
\label{eq:linear_stability}
\end{equation}
where ${\cal L}_{\rm v}$ is defined in eq.~\eqref{eq:Lv}, and 
\begin{equation}
\begin{split}
	\bar{\mathcal{L}}_\mathrm{f} &= \partial_v^2 +  (\partial_v - \alpha \partial_x ) v ,
	\\
	\alpha &= \sqrt{\tau} - 2 \rho \tau^{3/2} \int_0^\infty (\Psi')^2 e^{-\Psi} \dd r .
\end{split} 
\label{eq:linear_operator}
\end{equation}
To analyze the time evolution of the perturbation given in eq.~\eqref{eq:linear_operator}, the difficulty lies in treating the effect of the operator ${\cal L}_{\rm v}$, which contains information about microscopic interactions. Then, it is convenient to expand $\varepsilon$ in the eigenfunctions of the bare theory, namely in the absence of ${\cal L}_{\rm v}$. Interestingly, as stated previously, the bare theory maps into underdamped passive Brownian motion, and one can readily find the solution in the literature~\cite{risken}; see also \cite{Bothe:2021}. The corresponding eigenfunctions, which we here call the Fourier-Hermite basis, are 
\begin{equation}
	\psi_{nk} ( x, v) = e^{ - i k ( x + \alpha v )} U_n ( v - 2 i \alpha k ) ,
\label{eq:basis}
\end{equation}
where $U_n$ is the Hermite function
\begin{equation}
	U_n (z ) =\frac{1}{\sqrt{2 \pi } } H_n ( z)\, e^{- z^2/2 } = \frac{(-1)^n}{\sqrt{2\pi}}\partial_z^n e^{-z^2/2} ,
\label{eq:hermite_def}
\end{equation}
with the property that $(\partial_z^2 + \partial_z z ) U_n(z) = -n U_n(z)$. Indeed, acting on the Fourier-Hermite basis with the modified free operator $\bar{\cal L}_{\rm f}$ yields
\begin{equation}
	\bar{\mathcal{L}}_\mathrm{f} \psi_{nk} = - \lambda_{kn} \psi_{nk} ,
	\quad
	\lambda_{kn} = (\alpha k)^2 + n .
\end{equation}
As the operator $\bar{\mathcal{L}}_\mathrm{f}$ is not Hermitian, the conjugate basis is not simply the complex conjugate. Instead, we introduce the following conjugate basis
\begin{equation}
\begin{split} 
	\bar{\psi}_{nk} ( x, v) &= e^{i k ( x + \alpha v )} \bar{U}_n(v - 2 i \alpha k ) ,
	\\
	\bar{U}_n(z ) &= \frac{1}{n!} H_n (z) ,
\end{split} 
\label{eq:dual_basis}
\end{equation}
yielding
\begin{equation}
	\int \dd x \dd v \,\bar{\psi}_{mk'} ( x, v ) \psi_{nk} ( x, v) = 2 \pi \delta(k - k') \delta_{mn} ,
\end{equation}
so that the orthogonality relation between the basis $\psi_{nk}$ and its conjugate $\bar\psi_{nk}$ indeed holds as expected.

Having obtained the eigenvectors and eigenvalues of the bare theory, we decompose the perturbation $\varepsilon$ in eigenfunctions of $\bar{\mathcal L}_{\rm f}$ as
\begin{equation}
	\varepsilon(x, v, t) = \sum_{n} \int \varepsilon_{kn}(t) \psi_{nk}(x, v) \mathrm{d} k .
\end{equation}
Multiplying eq.~\eqref{eq:linear_stability} with the conjugate basis $\bar{\psi}_{kn}$ and integrating over $\{x, v\}$ yields the dynamics of the perturbation $\varepsilon$, in the Fourier-Hermite basis, as
\begin{equation}
	\dot\varepsilon_{km} = - \lambda_{km} \varepsilon_{km} + \sum_n \Big[ M^{(1)}_{kmn} + M^{(2)}_{kmn} + M^{(3)}_{kmn} \Big] \varepsilon_{kn} ,
\end{equation}
where
\begin{equation}
\begin{split} 
	M^{(1)}_{kmn} &= 2 \tau \rho {\cal L}_{{\rm a},k}  \frac{(-i \alpha k)^{n+m}}{ \alpha m!} e^{(\alpha k)^2} (m-n) ,
	\\
	M^{(2)}_{kmn} &=  - 2 \tau \rho {\cal L}_{{\rm b},k} k^2 \frac{(-i\alpha k)^{n+m}}{m!} e^{(\alpha k)^2} ,
	\\
	M^{(3)}_{kmn} &= \tau \rho {\cal L}_{{\rm c},k} \frac{(-i\alpha k)^{n+m}}{\alpha^2 m!} e^{(\alpha k)^2}
	\\
	& \quad\times \Big[ m(m-1) + n(n-1) - 2 mn -2(\alpha k)^2 \Big] .
\end{split} 
\end{equation}
Here, ${\cal L}_{{\rm a},k} = \int e^{ikx} {\cal L}_{\rm a} \dd k$, with similar definitions for ${\cal L}_{{\rm b},k}$ and ${\cal L}_{{\rm c},k}$, where the operators $\{{\cal L}_{\rm a}, {\cal L}_{\rm b}, {\cal L}_{\rm c}\}$ are defined in eq.~\eqref{eq:Labc}. If the matrix $ M_{knm} = - \lambda_{km} \delta_{mn} + M^{(1)}_{kmn} + M^{(2)}_{kmn} + M^{(3)}_{kmn}$, which controls the growth rate of the perturbation $\varepsilon$, has no positive eigenvalue, the uniform solution is stable; otherwise the system undergoes spinodal decomposition. Hence we only need the largest eigenvalue of $M$. As shown in the SM, the matrix $M$ can be diagnonalized exactly. This calculation actually only amounts to finding the maximum eigenvalue of a $3 \times 3$ matrix, which can be done straightforwardly, while also perturbatively keeping track of orders of $\tau$.


\begin{figure} 
\includegraphics[width=0.49\textwidth]{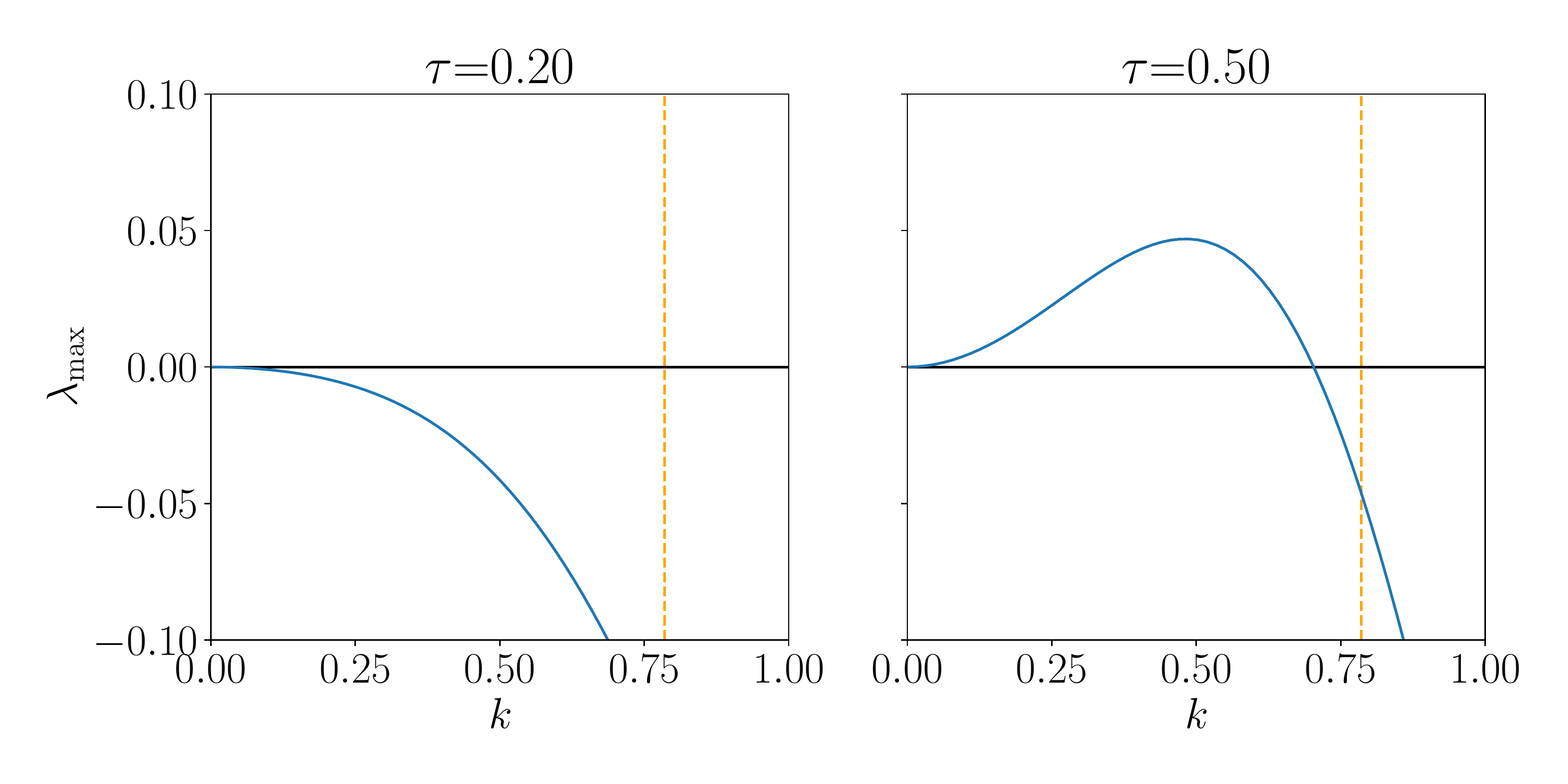}
\caption{The value of the largest eigenvalue $\lambda_\mathrm{max}$ as a function of the wavevector $k$ for $\nu = 10, r_0 = 2, \phi=0.6$, varying $\tau$(blue solid line). The orange dotted line represents the short length-scale cut-off as we are only concerned with length-scales beyond the particle size $r_0$.  }
\label{fig:spinodal}
\end{figure}

\begin{figure} 
\centering
\includegraphics[width=0.49\textwidth]{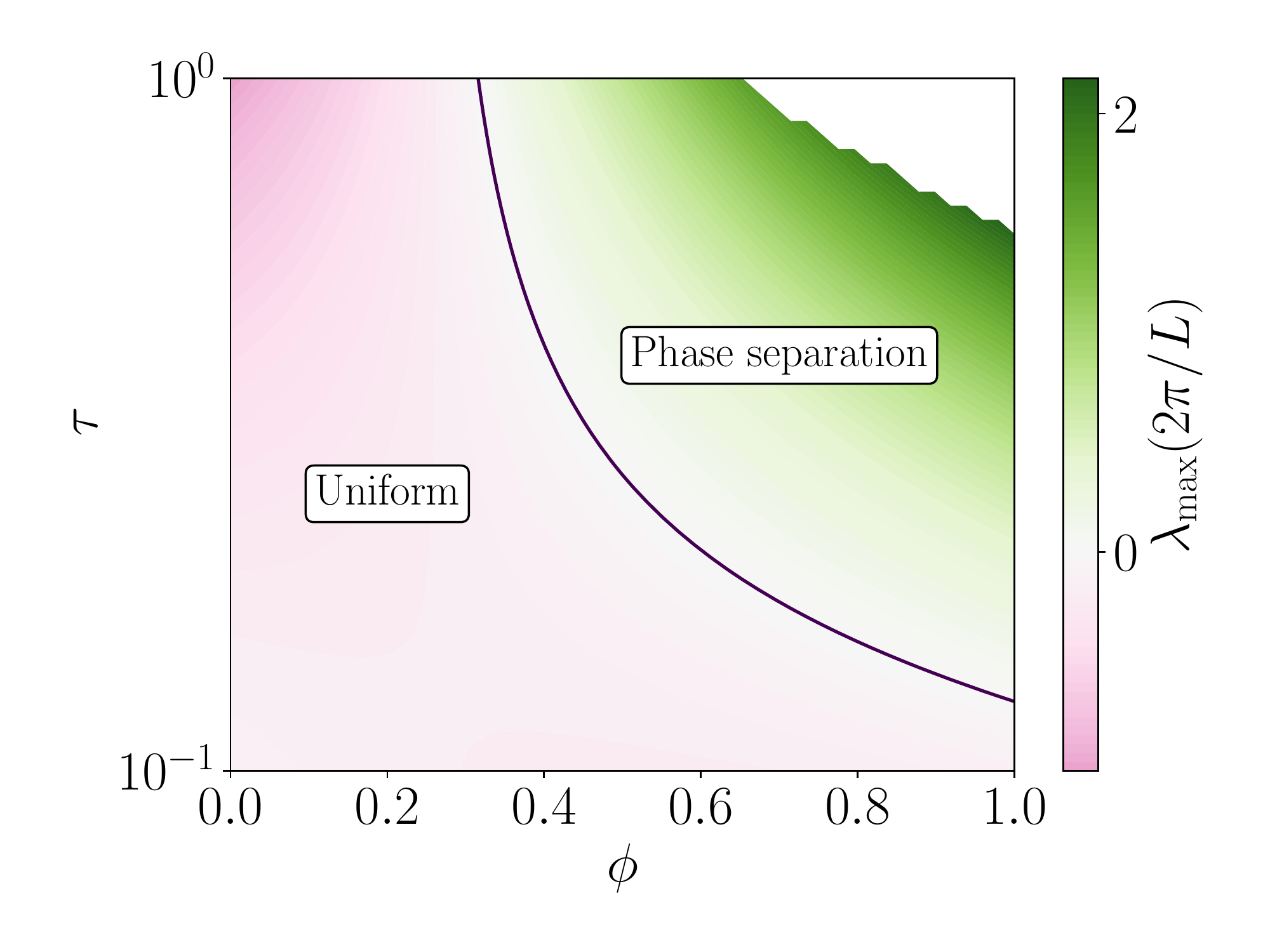}
\caption{Stability diagram in $\tau-\phi$ space showing $\lambda_\mathrm{max} (2 \pi / L)$ for $L = 200, \nu = 10, r_0 = 2$. The white region in the top-right corner corresponds to when $\alpha < 0$, where the method breaks down, as it is no longer in the small $\tau$ limit. The solid black line is the spinodal instability.}
\label{fig:phase_diagram}
\end{figure}

{\bf Spinodal instability}~--~We now illustrate how our approach, valid for an arbitrary pairwise potential, can be deployed to detect the onset of instability. We consider the following short-ranged, weakly repusive interaction potential: 
\begin{equation}
\Psi (r ) = \nu \exp ( - \frac{1}{1 - (r /r_0 )^2 } ) ,
\end{equation}
characterised by its strength $\nu$ and range $r_0$, whose derivatives are continuous at any order for $r$ within $[0,r_0]$. The maximum eigenvalue of the corresponding growth rate matrix $M$, expanded analytically to first order in the persistence time $\tau$, is plotted against wavevector $k$ in Fig.~\ref{fig:spinodal} for various values of $\tau$ and volume fraction $\phi = 2 r_0 N/V$, where $2 r_0$ is taken as the size of the particle in 1D and $V=L$ is the system volume. As expected~\cite{mips}, the system exhibits a spinodal instability at high $\tau$ and packing fraction $\phi$, for which the linear perturbation $\varepsilon$ is unstable, namely $\lim_{k \to 0} \lambda_\mathrm{max} (k ) = 0^+$. Fig.~\ref{fig:phase_diagram} shows the stability diagram based on the sign of the longest wavelength perturbation, $\lambda_\mathrm{max} ( 2 \pi / L ) $, as a function of $\tau$ and $\phi$, showing the transition between a uniform stable and a phase-separated state. (Note that we could equally have chosen the fastest growing mode to locate the spinodal.)

It is notable that our theory of interacting AOUPs shows a spinodal instability. As in equilibrium systems with attractions, it does this even at lowest order in the virial-type expansion in powers of density. Just as holds there, finding the instability requires using a low-density theory at finite densities. However the purpose of this approach is not to gain quantitative predictions about the exact location of the spinodal curve, but rather to establish that the macroscopic conditions for phase separation can be satisfied, by considering microscopic interaction laws and dynamics. Notably, unlike in equilibrium, the spinodal instability happens here for purely repulsive interactions -- a key feature of MIPS~\cite{mips}. Our microscopic calculation complements previous viewpoints based on effective quasi-equilibrium attractions~\cite{Farage2015}, and/or collisional slowing down~\cite{stenhammarPRL, Brady2015}.

Nonetheless, there are several caveats and limitations to our method. Firstly, the chosen potential is not strictly hard-core, but depends on the strength parameter $\nu$. In fact it is not possible to implement a true hard-core potential due to the nature of the small $\tau$ expansion for the two-particle density, as there are terms directly proportional to $\Psi$ or its derivatives. Unsurprisingly then, the results presented here do depend on $\nu$. While overly large $\nu$ will break the small-$\tau$ expansion, we have checked that a range of moderate values produce stability diagrams qualitatively similar to Fig.~\ref{fig:phase_diagram}. Secondly, for some parameters, we found that the largest eigenvalue stays negative  at low wavenumber but turns positive (unstable) at large ones, resembling an upside-down version of the right frame of Fig.~\ref{fig:spinodal}. At first sight this might be taken as evidence of new physics in the form of microphase separation at finite wavenumber, an outcome generically predicted by some field-theoretic models~\cite{tjhung:2018}. However, when this happens the wavenumbers in question (at or around the orange line in Fig.~\ref{fig:spinodal}) are comparable to the inverse particle separation $1/r_0$.  As the theory proposed here is a macroscopic one, it may not operate reliably in this large $k$ region, so we do not take this as evidence of microphase separation in repulsive AOUPs.

{\bf Conclusion}~--~Inspired by liquid-state theory of equilibrium systems, we have started with the hierarchy of density correlations for AOUPs, and closed it at second order using a quasistatic approximation. This approach is closely related to the virial expansion in the equilibrium case. It yields a mean field theory for the one-point density from first principles. We have analysed the stability of the uniform solution by looking at the largest eigenvalue of the growth matrix, and determined the onset of a spinodal instability. 

Our method deals directly with particle velocities as well as positions. Perhaps more importantly, it is able to capture (to lowest order in density) the physics of steep, short-ranged interactions. In that respect, it is distinct from standard coarse-graining methods for non-equilibrium systems, such as those based on Dean's equation~\cite{dean}, which starts as an exact representation but upon assuming a smooth (coarse-grained) density becomes an expansion in weak or slowly varying interaction forces. The Dean's approach neglects strong, short-range interactions that change the statistics of close encounters even when the density, and hence the {\em average} interaction force, is small. In equilibrium statistical mechanics, the virial expansion is designed to handle exactly this situation. We have presented the leading order counterpart of this, for an active system comprising interacting AOUPs.

There are some limitations to our method. Firstly, the two-particle stationary solution used in the quasistatic ansatz is not exact, but rather a small-persistence approximation, unlike the equilibrium case. Secondly, in truncating the density-correlation hierarchy at the second order, we assumed that the number density is low. Hence, even assuming small persistence, our approach is quantitatively accurate only in the dilute limit. However, it is a triumph of equilibrium liquid-state theories that the small-density virial expansion successfully predicts liquid-gas phase transitions, even though the regimes of phase separation are often well beyond the dilute limit. Here, we appeal to similar arguments to claim that our method gives bottom-up confirmation of how phase separation can occur in purely repulsive active systems, and it can be used to study how the spinodal density varies with interaction parameters and with the persistence time. 

Going forward, as we alluded to previously, the fact that the two-point velocity correlation is not a simple product of single particle distributions has profound consequences for other thermodynamic quantities, such as pressure \cite{pressure, Brady2014, Marchetti2014} and dissipation \cite{dissipation, Nemoto2020}. It would be interesting to explore the implications of the velocity correlations by calculating such quantities from the two-particle ansatz directly. Secondly, our approach here is limited to soft potentials and small persistence due to the reliance on the small $\tau$ expansion of the two particle distribution. Suppose there were an active system that is exactly solvable for two particles, or if one were to supplement the calculation with numerical simulations of the two-particle dynamics, it would be possible to obtain the phase separating dynamics for high persistence and hard-core potentials via our method. 

\acknowledgements{The authors acknowledge insightful discussions with Alexander Grosberg, Jean-Francois Joanny and Marius Bothe. YIL acknowledges support from Royal Society grant (RP\textbackslash R1\textbackslash 180165). \'EF acknowledges support from the Luxembourg National Research Fund (FNR), grant reference 14389168. Work funded in part by the European Research Council under the EU's Horizon 2020 Programme (Grant number 740269). RG-M acknowledges support from a St John's College Research Fellowship, University of Cambridge.}


\bibliographystyle{eplbib}
\bibliography{References}

\end{document}


\maketitle


\section{Overdamped passive Brownian particles: Quasistatic ansatz for density correlations} 

In this section, we demonstrate that, for overdamped passive Brownian particles (PBPs), one recovers the virial expansion of the free energy when using the quasistatic ansatz $p_2(\bx_1, \bx_2, t) = p_1 (\bx_1, t) p_1 (\bx_2, t) p_\mathrm{ss} (\bx_1, \bx_2 )$.  We consider the following equations of motion for $N$ PBPs with pair-interaction potential $\Psi$:
\begin{equation}
	\dot \bx_i = - \partial_{\bx_i} U  +  \sqrt{ 2 T } \bLambda_i , 
	\quad
	U = \sum_{i=1}^N \sum_{j<i} \Psi (| \bx_i - \bx_j |) , 
\end{equation}
where we have set the mobility to unity, so that $T$ is the diffusion constant, and $\bLambda_i$ is a Gaussian white noise. Similar to the main text, we write down the corresponding hierarchy for one- and two-particle densities, assuming that the density is sufficiently low to ignore contributions from three-particle density:
\begin{equation}
\begin{split}
	\partial_t p_1 (\bx_1, t ) &= T \nabla_{\bx_1}^2 p_1 (\bx_1, t) + \nabla_{\bx_1} \cdot \int p_2 (\bx_1, \bx_2, t ) \nabla_{\bx_1} \Psi (r) \dd \bx_2 ,
	\\
	\partial_t p_2 ( \bx_1, \bx_2, t ) &= T \left (  \nabla_{\bx_1}^2 + \nabla_{\bx_2}^2 \right ) \, p_2 (\bx_1, \bx_2, t )   + \sum_{i=1, 2}\nabla_{\bx_i} \cdot ( p_2 (\bx_1, \bx_2, t ) \nabla_{\bx_i} \Psi (r) ) ,
\end{split}
\end{equation}
where $r = |\bx_1 - \bx_2|$. In the quasistatic limit, where the interaction time is much faster than the free particle travel time, it is reasonable to suppose that every pair of particles reach steady state. This inspires the following ansatz to close the density hierarchy~\cite{GrosbergPRE, IlkerPRR}: 
\begin{equation}
	p_2 ( \bx_1, \bx_2, t ) = p_1 ( \bx_1, t) \, p_1 ( \bx_2, t ) \exp ( - \Psi (r) / T ),
\end{equation}
where we have used that the Boltzmann distribution is the the steady-state distribution of two particles. Substituting this ansatz into the $p_1$ equation, we have
\begin{equation}
	\partial_t p_1 (\bx_1, t) = T \nabla_{\bx_1}^2 p_1 (\bx_1, t) + \nabla_{\bx_1}  \cdot \left [ p_1 (\bx_1, t )  \int p_1 ( \bx_2, t ) e^{- \Psi (r) / T} \nabla_{\bx_1} \Psi ( r ) \dd \bx_2  \right ].
\label{eq:p1}
\end{equation}
Notice that $ e^{- \Psi (r) / T} \nabla_{\bx_1} \Psi ( r )  = - T \nabla_{\bx_1} f(r) $, where $f(r) =  \exp ( - \Psi ( r) / T ) - 1$ is usually called the Mayer-$f$ function~\cite{KardarParticles}. Changing variables from $\bx_2$ to $\br = \bx_1 - \bx_2$ in the integrand of eq.~\eqref{eq:p1}, and integrating by parts, we get
\begin{equation}
	\frac{1}{T} \partial_t p_1 ( \bx_1, t) = \nabla_{\bx_1}^2  p_1 ( \bx_1, t ) - \nabla_{\bx_1} \cdot \left [  p_1 (\bx_1,t) \int f(r) \nabla_{\bx_1} p_1 ( \bx_1 - \br,t) \dd \br  \right ] .
\label{eq:underdamped_mft}
\end{equation}
Thus far, we have closed the density hierarchy and obtained a mean-field equation for the one-particle density. In the following, we seek to connect eq.~\eqref{eq:underdamped_mft} with existing knowledge of equilibrium physics of interacting gases, in particular, the virial expansion.


\subsection{Virial expansion} 

We assume that the density distribution $\rho$ does not vary dramatically over the length-scale of the pair-potential. Therefore, expanding $p_1(\bx_1 - \br)$ in eq.~\eqref{eq:underdamped_mft} around $\bx_1$ yields
\begin{equation}
	\partial_t p_1 = \nabla \cdot ( p_1 \nabla \mu ) ,
	\quad
	\mu =  T (\log p_1 + 2 B p_1) ,
	\quad
	B = -\frac{1}{2} \int f (r) \dd \br .
\label{eq:virial}
\end{equation}
Here $B(T)$ is the second virial coefficient in equilibrium statistical physics~\cite{KardarParticles}. The stationary solution minimises the following free energy:
\begin{equation}
	\mathcal{F} = T \int \Big[ p_1 \log p_1 + B p_1^2 \Big] \dd \bx .
\end{equation}
The lack of gradient terms in the free energy is a result of neglecting higher-order terms in the expansion of $p_1 (\bx_1 - \br)$. Assuming that the stationary state is uniform with density $\rho = N/V$, we obtain 
\begin{equation}
\mathcal{F}( N, V, T) =  \underbrace{T N \log ( N/V)}_{\text{ideal gas}} + \underbrace{T B N^2/ V}_{\text{virial correction}} .
\end{equation}
This is exactly the equilibrium virial expansion to second order~\cite{KardarParticles}.


\subsection{All-gradient expansion} 

To explore the effects of higher-gradient terms, we expand $p_1(\bx - \br)$ in eq.~\eqref{eq:underdamped_mft} to all orders in gradient:
\begin{equation}
	p_1(\bx - \br ) = e^{-\br \cdot \nabla_\bx } p_1 (\bx) ,
\end{equation}
where $e^{-\br \cdot \nabla_\bx} = \sum_n \frac{(- \br \cdot \nabla_\bx )^n}{n!}$. Substituting into eq.~\eqref{eq:underdamped_mft}, we obtain
\begin{equation}
	\frac{1}{T} \partial_t p_1 = \nabla^2 p_1 + 2 \nabla \cdot \left [ p_1 \nabla \left ( \bar B p_1  \right ) \right ],
	\quad
	\bar B = - \frac{1}{2} \int f(r) e^{-\br \cdot \nabla_\bx} \dd \br ,
\end{equation}
where the second virial coefficient has the same form as in eq.~\eqref{eq:virial} but $\bar B$ is now an operator. Proceeding as in the previous section, the chemical potential $\mu$ has the same form as eq.~\eqref{eq:virial}. In Fourier space, $\bar B_k = \int e^{i\bk\cdot\bx} \bar B \dd\bx$ is the Fourier transform of the Meyer-$f$ function. Note that $\bar B_{k=0} = B$, since we only captured the zeroth mode when truncating the gradient expansion at the lowest order. The corresponding free energy is
\begin{equation}
	\bar {\cal F} = \underbrace{T \int p_1 \log p_1 \dd \bx}_{\text{ideal gas}} + \underbrace{T \int p_{1,k} \,\bar B_k \,p_{1,-k} \frac{ \dd \bk }{ (2 \pi)^d }}_{\text{interactions}} .
\end{equation}
Overall, the gradient expansion is a natural extension of the equilibrium virial expansion for non-uniform densities. We note that one cannot trust the high-$k$ behaviour of $B(\bk)$, as it is sensitive to the details of the potential at close range. However, since we are always interested in length-scales much larger than the range of the potential (i.e. size of the individual particles), we do not need to be concerned with such a high-$k$ behaviour.


\section{Steady state of AOUPs: Small-$\tau$ expansion}

In this section, we derive the stationary probability as an expansion at small $\tau$ . As opposed to~\cite{FodorPRL}, where the expansion was performed in the $(x, \dot{x})$ space, we consider here expansion in $(x, v)$ space. This is necessary as we need the steady-state two-particle distribution in $(x, v)$ space in the quasistatic ansatz. We start from the Fokker-Planck equation for two particles in one spatial dimension:
\begin{equation}
	\partial_t P = \sum_{i=1, 2} \mathcal{L}_i P ,
	\quad 
	\mathcal{L}_i = \frac{D}{\tau^2} \partial_{v_i}^2  + \left (  \frac{1}{\tau} \partial_{v_i} - \partial_{x_i} \right ) v_i + \partial_{x_i} ( \partial_{x_i} \Psi ) .
\end{equation}
Scaling units as $v \to v \sqrt{ \tau/ D}$, $x\to x/\sqrt{\tau D}$, $t\to t/\tau$, and $\Psi \to \Psi / D$, we obtain
\begin{equation}
	\mathcal{L}_i = \mathcal{L}_{i,0} + \sqrt{\tau} \mathcal{L}_{i,1} + \tau \mathcal{L}_{i,2} ,
\end{equation}
where
\begin{equation}
	\mathcal{L}_{i,0} = \partial_{v_i}^2 + \partial_{v_i} v_i ,
	\quad
	\mathcal{L}_{i,1} = - \partial_{x_i} v_i ,
	\quad
	\mathcal{L}_{i,2} = \partial_{x_i} ( \partial_{x_i}  \Psi ) .
\end{equation}
We assume the following $\tau$-expansion for the stationary distribution:
\begin{equation}
	p_\mathrm{2, ss} = e^{-\frac{1}{2} \left ( v_1^2 + v_2^2 \right ) - \Psi } \left [ 1 + \sum_n (\sqrt{\tau} )^n A_i ( \{ x_i, v_i \} ) \right ],
\end{equation}
where $A_i$ is the $i$-th order term. Solving order by order, we get
\begin{equation} 
\begin{split} 
 p_\mathrm{2, ss} &\propto \exp [ -\Psi(r) - \frac{1}{2} \left ( v_1^2 + v_2^2  \right ) ] \\ 
 &\quad \times \left ( 1 + \sqrt{\tau} (v_1-v_2) \Psi'  + \tau \left [ \frac{(v_1-v_2)^2}{2} (  ( \Psi')^2 - \Psi '' ) - 2 (\Psi' )^2 + 3 \Psi ''  \right ] + o(\tau) \right ) ,
 \end{split} 
\end{equation}
where $\Psi' = d\Psi/dr$. This result is used to derive the quasistatic approximation in eq.~(12) of the main text.


\section{The interaction matrix in Fourier-Hermite basis} 

In this section, we show the details of the change of basis to the Fourier-Hermite basis. Recall that the perturbation density field $\varepsilon$ can be decomposed in the Fourier-Hermite basis as
\begin{equation}
	\varepsilon (x, v) = \sum_{n} \int \varepsilon_{kn}(t) \psi_{nk}(x, v) \mathrm{d} k ,
	\quad
	\psi_{nk} = e^{-ik(x+\alpha v)} U_n ( v - 2 i \alpha k ) ,
\end{equation} 
where the Hermite function $U_n$ is defined in eq.~(22) of the main text. Substituting the decomposition into eq.~(19) in the main text, we get
\begin{equation}
\begin{split} 
	\sum_{n} \int (\dot \varepsilon_{kn} + \varepsilon_{kn} \lambda_{nk}) \psi_{nk} (x, v) \mathrm{d} k &= - \tau \rho \sum_{n} \int i k \varepsilon_{kn} U_0(v) \mathrm{d} k \int \psi_{nk} ( x, v-w) \dd w
	\\
	&\quad\times\Big[ 2 {\cal L}_{{\rm a},k} w + 2 {\cal L}_{{\rm b},k} (-ik) + {\cal L}_{{\rm c},k} (-ik) w^2  \Big] ,
\end{split}
\label{eq:eigen}
\end{equation}
where
\begin{equation}
\begin{split}
	{\cal L}_{{\rm a},k} &= \sqrt{\tau}  \int_0^\infty \mathrm{d} r \, ( \Psi')^2 \, e^{-\Psi} e^{ i r k} ,
	\\
	{\cal L}_{{\rm b},k} &= - \int_0^\infty \dd r f_0 ( r) e^{i  r k}, \quad f_0 (s) = \int_s^\infty \dd r \, \Psi' e^{-\Psi} \left [ 1 -  \tau 2 ( \Psi' )^2  + \tau 3  \Psi'' \right ] ,
	\\
	{\cal L}_{{\rm c},k} &= - \int_0^\infty \dd r \, f_1(r)   e^{i r k}, \quad f_1(s) = \tau \int_s^\infty \dd r \Psi' ( (\Psi' )^2 - \Psi'' ) e^{-\Psi} .
\end{split} 
\end{equation}
We then multiply both sides of eq.~\eqref{eq:eigen} by the dual basis $\bar{\psi}_{mk'}$, as defined in eq.~(24) of the main text, and integrate over the $(x, v)$. By orthogonality of the basis, as outlined in eq.~(25) of the main text, we have 
\begin{equation}
	\dot \varepsilon_{k m} = - \lambda_{k m }\varepsilon_{k m} +  \sum_n \Big[ M^{(1)}_{kmn} + M^{(2)}_{kmn} + M^{(3)}_{kmn} \Big] \varepsilon_{k n} ,
\end{equation}
where
\begin{equation}
\begin{split}
	M^{(1)}_{kmn} & = 2 \tau\rho {\cal L}_{{\rm a},k} ( -ik)  \int \dd v \, U_0(v) \bar{U}_m (v-2i \alpha k) \int \dd w \, w \, e^{i \alpha kw}  U_n (v-w-2i \alpha k) ,
	\\
	M^{(2)}_{kmn} &= 2 \tau\rho {\cal L}_{{\rm b},k} (-ik)^2  \int \dd v \, U_0(v) \bar{U}_m (v-2i \alpha k) \int \dd w \,  e^{i \alpha kw}  U_n (v-w-2i \alpha k) ,
	\\
	M^{(3)}_{kmn} &= \tau\rho {\cal L}_{{\rm c},k} (-ik)^2 \int \dd v \, U_0(v) \bar{U}_m (v-2i \alpha k) \int \dd w \, w^2  e^{i \alpha kw}  U_n (v-w-2i \alpha k) .
\end{split} 
\end{equation}
We first perform all the $w$-integrals:
\begin{equation}
\begin{split} 
	I_0^n (v, k) &\equiv \int \dd w \, e^{ik \alpha w} U_n(v-w-2i \alpha k) = e^{\frac{3}{2} \alpha^2 k^2 + i \alpha k v } (-i\alpha k)^{n}
	\\
	I_1^n(v, k) &\equiv \int \frac{\dd w}{\sqrt{2 \pi } }  \, w \, e^{i \alpha kw} U_n(v-w-2i \alpha k) = - e^{\frac{3}{2} \alpha^2 k^2 + i \alpha k v } (-i \alpha k)^{n-1} \left [ i \alpha k v + n + \alpha^2 k^2 \right ]
  \\
	I_2^n ( v, k ) &\equiv \int \frac{\dd w}{\sqrt{2 \pi } }  \, w^2 e^{i \alpha kw} U_n(v-w-2i \alpha k)
	\\
	&= e^{\frac{3}{2} \alpha^2 k^2 + i \alpha k v } (-i \alpha k)^{n-2} \left [ n(n-1)  - \alpha^2 k^2 ( 1 + (v - i \alpha k )^2 ) + 2 i n  \alpha k (v  - i \alpha k ) \right ] .
\end{split}
\end{equation}
Performing next the $v$-integrals, we find:
\begin{equation}
\begin{split} 
	M^{(1)}_{kmn} &= 2 \tau\rho {\cal L}_{{\rm a},k} (-ik) \int \frac{\dd v}{\sqrt{2\pi}} e^{-\frac{1}{2} v^2} H_m(v-2i \alpha k) I_1^n(v, k)
	\\
	&= 2 \tau\rho {\cal L}_{{\rm a},k} \frac{(-i \alpha k)^{n+m} }{ \alpha m!} e^{ \alpha^2 k^2} (m-n) ,
	\\
	M^{(2)}_{kmn} &= - 2 \tau\rho {\cal L}_{{\rm b},k} k^2 \int \frac{\dd v}{\sqrt{2\pi}} e^{-\frac{1}{2} v^2} H_m(v-2i \alpha k) I_0^n(v, k)
	\\
	&= - 2 \tau\rho {\cal L}_{{\rm b},k} k^2 \frac{(-i \alpha k)^{n+m} }{m!} e^{ \alpha^2 k^2} ,
	\\
	M^{(3)}_{kmn} &= \tau\rho {\cal L}_{{\rm c},k} (-ik)^2 \int \frac{\dd v}{\sqrt{2\pi}} e^{-\frac{1}{2} v^2} H_m(v-2i \alpha k) I_2^n(v, k) \\
	&= \tau\rho {\cal L}_{{\rm c},k} \frac{(-i \alpha k)^{n+m} }{ \alpha^2 m!} e^{ \alpha^2 k^2} \left [ m(m-1) + n(n-1) - 2 mn -2 \alpha^2 k^2 \right ] .
\end{split} 
\end{equation}
If the overall growth rate matrix $M_{kmn} = - \lambda_{km} \delta_{mn} + M^{(1)}_{kmn} + M^{(2)}_{kmn} + M^{(3)}_{kmn}$ has no positive eigenvalue, the uniform solution is stable; otherwise the system undergoes spinodal decomposition. Hence, all we need is the sign of the largest eigenvalue of the $M$ matrix.

Despite the appearance of $M$ as an infinite-dimensional matrix, it is in fact possible to find another set of basis where $M$ is block diagonal. Indeed, $\{M^{(1)}, M^{(2)}, M^{(3)}\}$ are all linear combinations of the following matrices:
\begin{equation}
\frac{(-i \alpha k)^{m+n}}{m!}, \, \frac{(-i \alpha k)^{m+n}}{m!} m, \, \frac{(-i \alpha k)^{m+n}}{m!} n, \, \frac{(-i \alpha k)^{m+n}}{m!} m(m-1), \, \frac{(-i \alpha k)^{m+n}}{m!} n(n-1) ,
\end{equation}
so that they can be written as
\begin{equation}
\begin{split}
	M^{(1)} &= \frac{2 \tau\rho {\cal L}_{{\rm a},k}}{\alpha}  e^{ \alpha^2 k^2} \left ( u_1 v_0^\intercal - u_0 v_1^\intercal \right ) ,
	\\
	M^{(2)} &= - 2 \tau\rho {\cal L}_{{\rm b},k} k^2  e^{ \alpha^2 k^2} u_0 v_0^\intercal ,
	\\
	M^{(3)} &= \frac{\tau\rho {\cal L}_{{\rm c},k}}{\alpha^2} e^{ \alpha^2 k^2} \left [ u_2 v_0^\intercal + u_0 v_2^\intercal - 2 u_1 v_1^\intercal - 2 \alpha^2 k^2 u_0 v_0^\intercal \right ] .
\end{split}
\end{equation}
where $\intercal$ denotes matrix transpose, and
\begin{equation}
\begin{split} 
	(u_0)_m &= \frac{(-i \alpha k)^{m}}{m!},
	\quad
	(u_1)_m = \frac{(-i \alpha k)^{m}}{m!} m,
	\quad
	(u_2)_m = \frac{(-i \alpha k)^{m}}{m!} m(m-1),
	\\
	(v_0)_n &= (-i \alpha k)^n,
	\quad
	(v_1)_n = (-i \alpha k)^n n,
	\quad
	(v_2)_n = (-i \alpha k)^n n (n-1),
\end{split} 
\end{equation}
Observe that when acting on an arbitrary vector $y$, $u v^\intercal y = ( v^\intercal y ) u$, meaning that a matrix of this form projects all vectors onto the direction of $u$. We introduce $\Delta = M^{(1)} + M^{(2)} + M^{(3)}$ in this section, and, grouping the terms, we have
\begin{equation}
	e^{-(\alpha k)^2} \Delta = h_{0} u_0 v_0^\intercal +h_{10} ( u_1 v_0^\intercal - u_0 v_1^\intercal ) + h_{11} u_1 v_1^\intercal + h_2 ( u_2 v_0^\intercal + u_0 v_2^\intercal ) ,
\end{equation}
where
\begin{equation}
	h_0 = - 2 k^2 \tau\rho \left ( {\cal L}_{{\rm b},k} + {\cal L}_{{\rm c},k} \right ) ,
	\quad
	h_{10} = 2 \tau\rho {\cal L}_{{\rm a},k} / \alpha ,
	\quad
	h_{11} = 2 \tau\rho {\cal L}_{{\rm c},k} / \alpha^2 ,
	\quad
	h_2 = \tau\rho {\cal L}_{{\rm c},k} / \alpha^2 .
\end{equation}
We now consider the following spanning set: $( v_0, v_1, v_2, ...  )$, where $(v_j)_n = ( - i \alpha k)^n P_{n,j}$ (recall $P_{n,j} = n!/(n- j)!$ is the permutation coefficient). One can easily check that it is a spanning set by spotting that the first $j$ elements of $v_j$ are zeros, and hence no vector in this set can be written as a linear combination of others. Furthermore, let us introduce $(q_0, q_1, q_2, .. )$ as the dual basis of $\{ v_j \}$, such that $v_i^\intercal q_j = \delta_{ij}$. We proceed to write $\Delta$ in the new basis: $\tilde{\Delta}_{ij} \equiv v_i^\intercal \Delta 
\, q_j$, where
\begin{equation}\label{eq:matrix}
\tilde{\Delta} = e^{ \alpha^2 k^2} 
\begin{pmatrix} 
h_0 v_0^\intercal u_0 + h_{10} v_0^\intercal u_1 + h_2 v_0^\intercal u_2
& -h_{10} v_0^\intercal u_0 + h_{11} v_0^\intercal u_1 & h_2 v_0^\intercal u_0 &  0 &  \dots \\ 
h_0 v_1^\intercal u_0 + h_{10} v_1^\intercal u_1 + h_2 v_1^\intercal u_2 
& -h_{10} v_1^\intercal u_0 + h_{11} v_1^\intercal u_1  & h_2 v_1^\intercal u_0 & 0 & \dots \\ 
h_0 v_2^\intercal u_0 + h_{10} v_2^\intercal u_1 + h_2 v_2^\intercal u_2 
& -h_{10} v_2^\intercal u_0 + h_{11} v_2^\intercal u_1 & h_2 v_2^\intercal u_0 & 0 & \dots \\
0 & 0 & 0 & 0 &  \dots \\
\vdots & \vdots & \vdots & \vdots  & \ddots 
\end{pmatrix},
\end{equation}
which is nonzero only in the top $3 \times 3$ due to the orthogonality relation between $v$'s and $q$'s. Computing the inner products and setting $\tilde{k} = \alpha k$ for convenience, we have
\begin{equation}
\begin{split} 
	v_0^\intercal u_0 &= \sum_n \frac{(- \tilde{k}^2)^n }{n!} = e^{-\tilde{k}^2} ,
	\\
	v_1^\intercal u_0 = v_0^\intercal u_1 &= \sum_n \frac{(- \tilde{k}^2)^n }{n!} n=- \tilde{k}^2 \sum_n \frac{(- \tilde{k}^2)^n }{n!} = - \tilde{k}^2 e^{-\tilde{k}^2} ,
	\\ 
	v_2^\intercal u_0  = v_0^\intercal u_2 &= \sum_n \frac{(-\tilde{k}^2)^n }{n!} n (n-1) = \tilde{k}^4 e^{-\tilde{k}^2} ,
	\\
	v_1^\intercal u_1 & =  \sum_n \frac{(-\tilde{k}^2)^n }{n!} \left [ n (n-1) + n \right ] = ( \tilde{k}^4 -\tilde{k}^2 ) e^{- \tilde{k}^2} ,
	\\
	v_1^\intercal u_2  = v_2^\intercal u_1 &= \sum_n \frac{(-\tilde{k}^2)^n }{n!} \left [ n (n-1) (n-2) + 2 n (n-1) \right ] = ( - \tilde{k}^6 + 2 \tilde{k}^4 ) e^{-\tilde{k}^2} ,
	\\
	v_2^\intercal u_2 &= \sum_n \frac{(-\tilde{k}^2)^n }{n!} \left [ n (n-1) (n-2) (n-3) + 4 n (n-1) (n-2) + 2 n(n-1)\right ] 
	\\
	&= ( \tilde{k}^8 - 4 \tilde{k}^6 + 2 \tilde{k}^4 ) e^{- \tilde{k}^2}.
\end{split} 
\end{equation}
The resulting $3\times 3$ matrix appearing in eq.~\eqref{eq:matrix} can then be written as
\begin{equation}
\begin{pmatrix} 
h_0 - h_{10} \tilde{k} ^2 + h_2 \tilde{k} ^4 & - h_{10} - h_{11} \tilde{k}^2 & h_2 \\ 
- h_0 \tilde{k}^2 + h_{10} \left ( \tilde{k}^4 -\tilde{k}^2 \right ) +  h_2 \left ( -\tilde{k}^6 + 2 \tilde{k}^4 \right )  & h_{10}  \tilde{k}^2 + h_{11} \left ( \tilde{k}^4- \tilde{k}^2 \right ) & - h_2 \tilde{k}^2  \\ 
h_0 \tilde{k}^4  + h_{10} \left (  -\tilde{k}^6 + 2 \tilde{k}^4 \right )  + h_2 \left ( \tilde{k}^8 - 4 \tilde{k}^6 + 2 \tilde{k}^4 \right )  & - h_{10} \tilde{k}^4 + h_{11} \left ( -\tilde{k}^6 + 2 \tilde{k}^4 \right )  & h_2 \tilde{k}^4 
\end{pmatrix}. 
\end{equation}
To compute the eigenvalues of the full matrix $M= D + \Delta$, where $D_{kmn} = (-n- \tilde{k}^2) \delta_{mn}$, we also need to write $D$ in the new basis:
\begin{equation}
	\tilde{D}_{jl} = v_j^\intercal D \, q_l = - \delta_{j+1, l} - (j - \tilde{k}^2) \delta_{jl} ,
\end{equation}
where we have used $n(v_j)_n = (v_{j+1} + j v_j)_n$, or explicitly
\begin{equation}
\tilde{D} = 
\begin{pmatrix} 
-\tilde{k}^2 & -1 & 0 & \dots \\
0 & -1 -\tilde{k}^2 & -1 & 0 &\dots \\ 
0 & 0 & - 2 - \tilde{k}^2 & -1 & 0 & \dots  \\
\vdots & \vdots & \vdots & \ddots & \ddots 
\end{pmatrix} .
\end{equation}
Notice that $\tilde{D}$ is upper triangular and therefore the eigenvalues are simply the diagonal elements $-n-\tilde{k}^2$, which are the same as for $D$.

Having written both the diagonal and non-diagonal parts in the new basis, we proceed to solve for the eigenvalues $\lambda$ of $\widetilde{M} = \tilde{D} + \tilde{\Delta}$, satisfying $\det ( \widetilde{M} - \lambda I ) =  0 $, where $I$ is the identity matrix. Observe that $\widetilde{M}$ is of the following form, 
\begin{equation}
\widetilde{M} = 
\begin{pmatrix} 
X & Y \\
0 & Z 
\end{pmatrix},
\end{equation}
where $X$ is a $3\times 3$ matrix. One can prove that $ \det ( \widetilde{M} - \lambda I ) = \det ( X - \lambda I ) \det ( Z - \lambda I )$. So we only need to solve for the eigenvalues of $X$ and $Z$ separately, but we already know the eigenvalues of $Z$ as it is upper-triangular with diagonal elements $(-n-\tilde{k}^2)$ for $n \geq 3$. Hence what remains are the eigenvalues of $X$, which can be solved numerically easily as it is only $3 \times 3$. Overall, we have reduced the problem of finding the maximum eigenvalue of an infinite-dimensional matrix to diagonalising the $3 \times 3$ matrix $\tilde{\Delta} + \tilde{D}$. We then compute the largest eigenvalue as
\begin{equation}
	\lambda_\mathrm{max} = - k^2 \bigg[ 1 + \tau \frac{1 + 2 (B_2(k) - 2 A_1(k)^2 + C_2(k)) - 4 k^2 B_0(k) (B_2(k) + C_2(k))}{1 - 2 B_0 (k)  k^2} \bigg] + o(\tau) ,
\end{equation}
where
\begin{equation}
	B_0(k) = {\cal L}_{{\rm b},k}\big|_{\tau=0} ,
	\quad
	B_2(k) = (d/d\tau) {\cal L}_{{\rm b},k} ,
	\quad
	A_1(k) = (1/\sqrt{\tau}) {\cal L}_{{\rm a},k},
	\quad
	C_2(k) = (1/\tau) {\cal L}_{{\rm c},k} .
\end{equation}


\bibliographystyle{eplbib}
\bibliography{SI}